\documentclass[
reprint,
superscriptaddress,
 amsmath,amssymb,
 aps,
pra,
]{revtex4-2}

\usepackage{graphicx}% Include figure files
\usepackage{subfigure}
\usepackage{dcolumn}% Align table columns on decimal point
\usepackage{bm}% bold math
\usepackage{xcolor}
\usepackage{caption}
\usepackage{float}
\usepackage{setspace}
\usepackage{booktabs}
\usepackage{amsmath} 
\usepackage{physics}
\usepackage{mhchem}
\usepackage[T1]{fontenc}
\usepackage{lmodern}
\usepackage{hyperref}

\hypersetup{
    colorlinks=true,      
    linkcolor=blue,       
    citecolor=green,      
    urlcolor=red,         
    pdfborder={0 0 0}     
}
\colorlet{nblue}{blue!65!cyan}
\colorlet{nred}{magenta!40!red}
\colorlet{ngreen}{green!50!nblue}

\newcommand{\partialsquare}[2][]{\frac{\partial^2 #1}{\partial #2^2}}

\newcommand{\Eq}[1]{Eq.~\eqref{#1}}
\newcommand{\eq}[1]{\eqref{#1}}
\newcommand{\Fig}[1]{Fig.~\ref{#1}}

\newcommand{\beq}{\begin{equation}}
\newcommand{\eeq}{\end{equation}}

\newcommand{\beqa}{\begin{eqnarray}}
\newcommand{\eeqa}{\end{eqnarray}}

\newcommand{\Beqa}{\begin{eqnarray*}}
\newcommand{\Eeqa}{\end{eqnarray*}}

\newcommand{\nn}{\nonumber}

\newcommand{\sign}{\text{sgn}}

\newcommand{\vect}[1]{\mathbf{#1}}
\newcommand{\I}{\mathrm{i}}

\begin{document}
\title{Scattering of a weakly bound dimer from a hard wall in one dimension}
%of one-dimensional mass-imbalanced dimers from a hard wall}
\author{Xican Zhang}
\author{Shina Tan}%
 \email{shinatan@pku.edu.cn}
\affiliation{%
International Center for Quantum Materials, Peking University, Beijing 100871, China 
}%
\affiliation{Beijing Key Laboratory of Quantum Devices, Peking University, Beijing 100871, China}
%\collaboration{CLEO Collaboration}%\noaffiliation
\date{\today}

\begin{abstract}
We consider a dimer formed by two particles with an attractive contact interaction in one dimension,
colliding with a hard wall. We compute the scattering phase shifts and the reflection coefficients for various collision energies and various mass ratios of
the two particles. For low-energy collisions (with dimer kinetic energies much smaller than the binding energy) our results are consistent with those of D. Lee and M. Pine, The European Physical Journal A \textbf{47}, 41 (2011).
For mass ratios much greater than 1 we use the Born-Oppenheimer approximation to show that the scattering length and the effective range of the dimer-wall collision both depend logarithmically on the mass ratio.
For collision energies much greater than the binding energy, the dissociation probability is inversely proportional to the square of the incident momentum of the dimer and we find the constant of proportionality analytically,
and we use a semiclassical analysis to approximately derive the ``angular distribution" of the dissociated pair,
where the ``angle" $\theta$ depends on the ratio of the velocities of the two outgoing unbound particles.
\end{abstract}

 \maketitle
\section{Introduction\label{sec:intro}}
The scattering of quantum particles from hard walls provides a fundamental paradigm for understanding boundary effects. In cold atom experiments, confined systems with tunable interactions can be created in tightly focused optical lattices \cite{PhysRevLett.81.938,RevModPhys.82.1225}. An additional laser may be focused to produce a sharp repulsive boundary \cite{Montangero_2009,Lee_2011}. Box traps can also be realized for ultracold atoms \cite{PhysRevLett.110.200406, van_es_box_2010, PhysRevLett.112.040403,PhysRevLett.118.123401,PhysRevLett.120.060402}, and few-body system in box traps have been studied \cite{olshanii_exactly_2015,PhysRevA.100.063623,LIU2019181}. The emergence of optical tweezer arrays enables the manipulation of individual molecules, such as the CaF molecule \cite{Anderegg_2019,PhysRevLett.125.043401} and the NaCs molecule \cite{PhysRevLett.126.123402,Liu_2018}. People have studied the collisions between one dimensional (1D) composite objects and obstacles  \cite{Li_2023,Moro_2010,Gomez_2019,Lee_2011}. The mass ratio between the constituent particles and the kinetic energy of the collision play important roles in the collision process. 

Recent experimental progress has enabled the preparation of ultracold heteronuclear molecules with a wide range of mass ratios~\cite{baroni_quantum_2024}. For example, mixtures of $^{40}\mathrm{K}$ and $^{6}\mathrm{Li}$ reach a mass ratio of approximately 6.7~\cite{LiK_1,LiK_2,LiK_3}, while combinations involving heavier fermionic isotopes such as $^{173}\mathrm{Yb}$ with $^{6}\mathrm{Li}$ yield mass ratios up to $\approx 29$~\cite{exp:YbLi_1,exp:YbLi_2}. Mass ratios exceeding these values are not currently achievable with stable atomic species. However, extreme mass ratios could be reached by tuning effective masses in optical lattices~\cite{effectivemassratios}. 
% For example, in a mixture of $^{133}\mathrm{Cs}$ and $^{6}\mathrm{Li}$, the effective mass ratio can be tuned to over 1000 by adjusting the lattice depths.

In this paper, we investigate the scattering of a weakly bound dimer from a hard wall in one dimension.  The dimer is formed by two distinguishable particles with masses $m_1$ and $m_2$ and with a short-range attractive interaction characterized by a $\delta$-function potential. We focus on two distinct scattering regimes: the low-energy regime, where the dimer reflects elastically, and the inelastic regime above the dissociation threshold, where the collision can break the dimer into its constituent particles.

\textbf{Model--} Let the coordinates of the two particles with masses $m_1$ and $m_2$ be $x_1$ and $x_2$ respectively. We assume that the particles interact with a delta function potential and that there is a hard wall at $x=0$. The two-body wave function $\Psi(x_1,x_2)$ for the energy eigenstate with energy $E$ satisfies the Schr\"{o}dinger equation
\begin{equation}\label{eq:se}
\Big[-\frac{1}{2 m_1}\partialsquare[]{x_1}-\frac{1}{2 m_2}\partialsquare[]{x_2}+g\delta(x_1-x_2)\Big]\Psi=E\Psi
\end{equation}
at $x_1,x_2>0$,
where we set $\hbar=1$, and the interaction strength $g=-1 /\mu a<0$  corresponds to an attractive interaction, $a$ is the one-dimensional $s$-wave scattering length between the particles, and $\mu = m_1 m_2 / (m_1+m_2)$ is the reduced mass. A hard wall is introduced at the origin by imposing the Dirichlet boundary conditions:
\begin{equation}
    \Psi(0,x_2)=\Psi(x_1,0)=0.
    \label{eq:dbc}
\end{equation}
The energy $E$ may be expressed as
\begin{equation}
    E=\frac{K^2}{2 M}-\frac{1}{2\mu a^2},
    \label{eq:E}
\end{equation}
where $K$ is the magnitude of the center-of-mass incident momentum, $M=m_1+m_2$ is the total mass of the dimer,
and the second term on the right hand side of the above equation is the energy of the bound state in its rest frame.
If we solve \Eq{eq:E} at $E=0$, we get a dissociation threshold momentum
\beq
K_\text{th}=\frac{m_1+m_2}{\sqrt{m_1 m_2}}\frac{1}{a}.
\eeq
We define the center-of-mass coordinate
\beq\label{X}
X=\frac{m_1x_1+m_2x_2}{m_1+m_2}
\eeq
and the relative coordinate
\beq\label{r}
r=x_2-x_1.
\eeq
For the dimer-wall collision, the wave function takes the following form
at $X\to\infty$:
\begin{equation}\label{Psi large X}
    \Psi(x_1,x_2)=e^{-|r|/a}\left[e^{-\I K X}+f e^{\I K X}\right]+\psi_d(x_1,x_2),
\end{equation}
where $f$ is the amplitude of the reflected wave, and $\psi_d(x_1,x_2)$ denotes the dissociated (unbound) part of the wave function which goes to zero at $X\to\infty$. The reflection coefficient is
\beq\label{R}
R=|f|^2.
\eeq
If $K<K_\text{th}$, the dimer scatters elastically off the hard wall, so that the reflection coefficient $R=1$ and the dissociation component $\psi_d(x_1,x_2)$ decays exponentially at $X\to\infty$. In this case, the wave function far from the wall reduces to a standing wave $\propto e^{-|r|/a}\sin(K X +\delta)$, where
\begin{equation}\label{delta}
\delta=\frac12\arg(-f)
\end{equation}
is the scattering phase shift. Here $\arg(z)$ means the argument of the complex number $z$.
If $K\ll K_\text{th}$ we have the effective range expansion,
\begin{equation}\label{kcotdelta}
    K \cot\delta=-\frac{1}{a_R}+\frac{1}{2}r_R K^2+\cdots,
\end{equation}
where $a_R$ is the dimer-wall scattering length, and $r_R$ is the dimer-wall effective range. If $K>K_\text{th}$, the incident dimer may dissociate after the collision, so that the reflection coefficient $R<1$. In this inelastic regime, 
the scattering phase shift $\delta$ is no longer defined in the standard sense. We extend the definition of the scattering phase shift to this regime by still using \Eq{delta}.

We solve our model with a combination of analytical and numerical methods.
Without loss of generality we assume that $m_1\ge m_2$ from now on.
In Sec.~\ref{sec: integrale cases} we derive exact solutions using the Bethe Ansatz 
for the special integrable cases of  $m_1/m_2=1$ \cite{Gaudin_1971} and $m_1/m_2=3$ \cite{LIU2019181}.
In Sec.~\ref{sec: general cases} we derive an integral equation for the problem for arbitrary mass ratios, and solve it numerically.
In Sec.~\ref{sec: BO}, for large mass imbalance ($m_1\gg m_2$), we employ the Born-Oppenheimer approximation to obtain analytical asymptotic results for $a_R$ and $r_R$.
In Sec.~\ref{sec: high energy}, for high-energy collisions with incident kinetic energies much larger than the binding energy of the dimer, we perform a semi-classical analysis and verify the validity of this analysis by comparing its predictions with those based on the numerical solutions in Sec.~\ref{sec: general cases}.

\section{Integrable cases \label{sec: integrale cases}}
Integrable systems are known to preserve chemical composition during scattering processes, even when the incident energy exceeds the dissociation threshold \cite{dodd_1972,PhysRevLett.96.163201}. 
The model introduced in Sec.~\ref{sec:intro} is integrable if the mass ratio $m_1/m_2$ is 1 or 3. The case $m_1/m_2=1$ was solved by Bethe ansatz by Gaudin~\cite{Gaudin_1971}. The ansatz for the mass ratio $m_1/m_2=3$ is based on a finite superposition of plane waves associated with a dihedral group $D_6$ as revealed in~\cite{LIU2019181}. This emergent symmetry is an example of the kinematic symmetries that arise in the ``no-tunneling'' limit imposed by an impenetrable barrier (the hard wall) as discussed in~\cite{Harshman_1, Harshman_2}.

If $m_1/m_2=1$,  the two-body wave function for the dimer colliding with the hard wall is
\begin{equation}
\begin{aligned}\label{Psi m1/m2=1}
    \Psi(x_1,x_2)&=e^{-\frac{|x_2-x_1|}{a}}\sin\left(K \frac{x_1+x_2}{2}+\delta\right)\\
    &\quad-e^{-\frac{x_1+x_2}{a}}\sin\left(K\frac{|x_1-x_2|}{2}+\delta\right),
\end{aligned}
\end{equation}
and the scattering phase shift $\delta$ satisfies \cite{Lee_2011,Gomez_2019}
\begin{equation}
    K \cot\delta=-\frac{2}{a}.
    \label{eq: delta1}
\end{equation}
Comparing \Eq{eq: delta1} with \Eq{kcotdelta}, one obtains $a_R=\frac{a}{2}$ \cite{Lee_2011} and $r_R=0$ \cite{Lee_2011}.

If $m_1/m_2=3$, 
% the integrability of the model is ensured by a dihedral $D_6$ symmetry \cite{LIU2019181}. 
setting $(k_1,k_2)=(-\frac34K+\frac{\I}{a},-\frac14K-\frac{\I}{a})$ and choosing the coefficients $A_{j\pm}$  in Eq.~(18) in Ref.~\cite{LIU2019181} to satisfy appropriate boundary conditions, we get the following exact wave function:
\begin{equation}\label{Psi m1/m2=3}
    \begin{aligned}
       \Psi(x_1,x_2)&= e^{-\frac{|x_1-x_2|}{a}}\sin\Big(K \frac{3x_1+x_2} {4}+\delta \Big) \\
       &\quad-e^{-\frac{x_1+x_2}{a}}\sin\Big[K \frac{3x_1-x_2}{4}\sign(x_1-x_2) +\delta \Big]\\
       &\quad+\frac{8}{\sqrt{16+9a^2K^2}}e^{-\frac{2 x_1}{a}} \sin\Big( \frac{K x_2}{2} \Big)\theta(x_1-x_2),
    \end{aligned}
\end{equation}
where $\sign(x)$ is the sign function (defined to be 1 for $x>0$ and $-1$ for $x<0$), $\theta(x)$ is the Heaviside step function, and the scattering phase shift $\delta$ satisfies
\begin{equation}
    K\cot\delta=-\frac{4}{3a}.
    \label{eq: delta3}
\end{equation}
Comparing \Eq{eq: delta3} with \Eq{kcotdelta}, we get $a_R=\frac34a$ and $r_R=0$.

Equations~\eqref{Psi m1/m2=1} and \eqref{Psi m1/m2=3} are valid for all $K>0$,
and thus the reflection coefficient $R=1$ for all $K$ (even if $K>K_\text{th}$) if $m_1/m_2$ is equal to $1$ or $3$.

\section{General Cases\label{sec: general cases}}
We begin by rewriting \Eq{eq:se} in a more convenient form,
\begin{equation}
\Big(\partialsquare[]{y_1}+\partialsquare[]{y_2}+2m_1E\Big)\Psi=-\frac{2(1+\beta^2)}{a}\delta(y_1-\beta y_2)\Psi,
\end{equation}
where we have introduced the rescaled coordinates,
\begin{equation}
    y_1=x_1, \quad y_2=\frac{x_2}{\beta}, \quad \text{with}\quad \beta=\sqrt{\frac{m_1}{m_2}}.
\end{equation}
Let $G(y_1,y_2,y_1',y_2')$ be the Green’s function for the noninteracting problem with hard-wall boundary conditions, satisfying
\begin{equation}\label{eq:se y coordinates}
    \Big(\partialsquare[]{y_1}+\partialsquare[]{y_2}+2m_1E\Big) G=\delta(y_1-y_1')\delta(y_2-y_2'),
\end{equation}
% The hard-wall boundary conditions \Eq{eq:dbc} translate to $\Psi=0$ at $y_1=0$ or $y_2=0$. 
with $G=0$ whenever $y_1=0$, $y_2=0$, $y_1'=0$ or $y_2'=0$. This Green’s function can be constructed via the method of images:
% Using the Green's function method, and assuming that the dissociated part $\psi_d(x_1,x_2)$ is an outgoing wave
% at large $X$ if $E>0$,
\begin{equation}
\begin{aligned}
     G(y_1,y_2,y_1',y_2')&=G_0\Big(\sqrt{(y_1-y_1')^2+(y_2-y_2')^2}\Big)\\&\quad+G_0\Big(\sqrt{(y_1+y_1')^2+(y_2+y_2')^2}\Big)\\&\quad-G_0\Big(\sqrt{(y_1-y_1')^2+(y_2+y_2')^2}\Big)\\&\quad-G_0\Big(\sqrt{(y_1+y_1')^2+(y_2-y_2')^2}\Big),
\end{aligned}
\end{equation}
and
\begin{equation}
    G_0(r)=\begin{cases}
    -\frac{1}{2\pi}K_0(\sqrt{-2m_1 E} \,r), & E<0\\
        \frac{1}{2\pi}\ln r, &E=0 \\
        -\frac{\I}{4}H^{(1)}_0(\sqrt{2m_1E}\, r),&E>0 \\
    \end{cases}.
\end{equation}
Here $K_0(\xi)$ is the decaying Bessel function, and $H_0^{(1)}(\xi)$ is the Hankel function of the first kind. Using the Green's function method, the wave function in scaled coordinates satisfies the Lippmann–Schwinger equation
\begin{equation}
    \Psi(\vect{y})=-\frac{2(1+\beta^2)}{a}\iint d^2\vect{y'}G(\vect{y},\vect{y'})\delta(y_1'-\beta y_2')\Psi(\vect{y'}),
\end{equation}
where $\vect{y}=(y_1,y_2)$, $\vect{y'}=(y_1',y_2')$, and the integral runs over the first quadrant, $y_1',y_2'>0$. Transforming back to the original coordinates $(x_1,x_2)$, we obtain
\begin{equation}\label{Psi}
    \Psi(x_1,x_2) = -\frac{2(1+\beta^2)}{a\beta}\int_0^{\infty} G(x_1,x_2/\beta,z,z/\beta)\psi(z)d z,
\end{equation}
where $\psi(z)\equiv\Psi(z,z)$. Setting $x_1=x_2=x$ in \Eq{Psi} yields a closed integral equation for
%$r=\sqrt{(y_1-z_1)^2+(y_2-z_2)^2}$. 
$\psi(x)$:
\begin{equation}\label{integral eq}
    \psi(x)=-\frac{2(1+\beta^2)}{a\beta}\int_0^\infty G(x,x/\beta, z,z/\beta)\psi(z)dz.
\end{equation}
According to \Eq{Psi large X}, at $x\to\infty$ we have
\beq\label{psi large x}
\psi(x)=e^{-\I Kx}+fe^{\I Kx}+\psi_d(x),
\eeq
where $\psi_d(x)\equiv\psi_d(x,x)$ goes to zero at $x\to\infty$.

If $E>0$, there is usually a nonzero probability $D$ for the dimer to dissociate after collision with the wall. 
For any fixed ratio $x_2/x_1\ne1$, we expand $\Psi$ in \Eq{Psi}
when $x_1$ and $x_2$ go to infinity simultaneously, and find that 
\begin{equation}
    \Psi(x_1,x_2)=c(\theta)\rho^{-1/2}e^{\I Q \rho}+o(\rho^{-1/2}),
\end{equation}
where $Q=\sqrt{2ME}$,
\begin{equation}\label{rho}
    \rho=\sqrt{\frac{m_1 x_1^2+m_2 x_2^2}{m_1+m_2}},
\end{equation}
\beq\label{theta}
\theta=\arctan\Big(\sqrt{\frac{m_2}{m_1}}\frac{x_2}{x_1}\Big),
\eeq

\begin{align}\label{c}
    c(\theta)&=e^{\I\frac{\pi}{4}}\frac{\sqrt2K_\text{th}}{\sqrt{\pi Q}} \int_0^\infty\big\{\cos\big[Q z\cos(\theta-\theta_0)\big]\nn\\
    &\qquad-\cos\big[Q z\cos(\theta+\theta_0)\big]\big\}\psi(z)dz,
\end{align}
and
\beq
\theta_0=\arctan\sqrt{m_2/m_1}.
\eeq
By studying the probability flux, we find that the probability that the two particles dissociate and
end up in the infinitesimal range $(\theta,\theta+d\theta)$ at large $\rho$ is $P(\theta)d\theta$
and
\begin{equation}
    P(\theta)=\frac{ Q K_\text{th}}{K}|c(\theta)|^2.
    \label{Ptheta}
\end{equation}
The total dissociation probability is
\begin{equation}
    D=\int_0^{\pi/2}P(\theta)d\theta.
    \label{D}
\end{equation}
One can analytically show that $R+D=1$.

\begin{figure}
    \centering
    \includegraphics[width=0.9\linewidth]{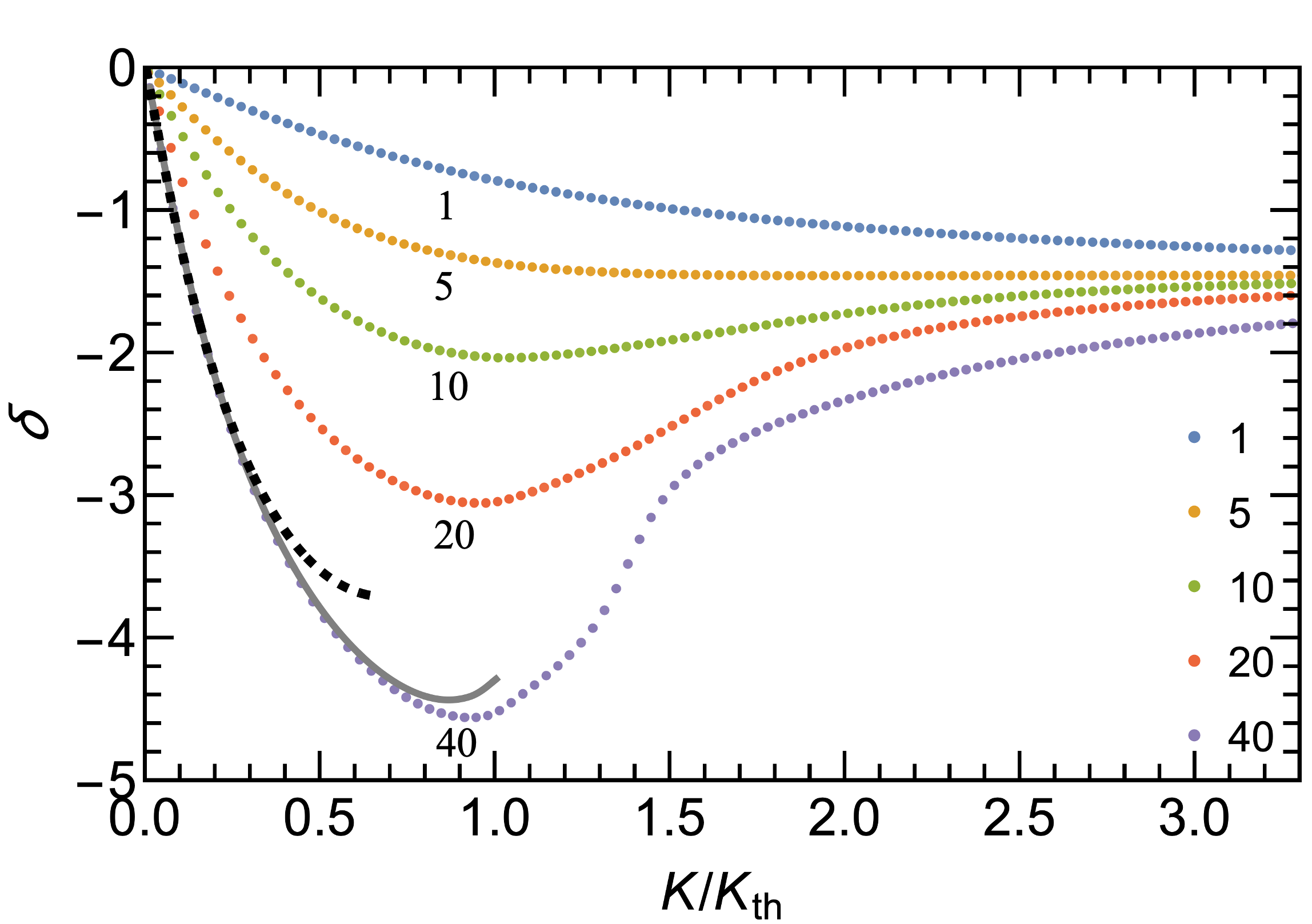}
    \caption{Dotted curves: the dimer-wall scattering phase shift $\delta$ versus $K/K_\text{th}$ for  mass ratios $1$, $5$, $10$, $20$, and $40$.  The numbers below the dotted curves are the corresponding mass ratios.
    Dashed curve: the prediction of \Eq{BO delta} based on an approximate treatment of the Born-Oppenheimer approximation for the mass ratio $m_1/m_2=40$.
    Grey curve:  the prediction of the Born-Oppenheimer approximation, based on the numerical solutions of \Eq{Schrodinger BO} together with Eqs.~\eqref{BO kappa exact} and \eqref{V eff} for $m_1/m_2=40$.}
    \label{fig:phaseshift}
\end{figure}

We have solved \Eq{integral eq} numerically to determine $f$ for various mass ratios and incident momenta.
We then computed the reflection coefficient $R$ and the scattering phase shift $\delta$, using Eqs.~\eqref{R} and \eqref{delta}.
For negative energies ($K<K_\text{th}$), the scattering is purely elastic and is fully characterized by the scattering phase shift $\delta$. %To maintain consistency across energy regimes, we employ the extended definition given in \Eq{delta} for all values of $K$.
For positive energies ($K>K_\text{th}$), the scattering is inelastic (unless $m_1/m_2$ equals 1 or 3), and the reflection coefficient $R<1$  (unless $m_1/m_2$ is 1 or 3).
%This extended definition reduces to the conventional phase shift in the elastic regime while providing a continuous and physically meaningful characterization of the scattering process even when $K>K_{th}$.
In Fig.~\ref{fig:phaseshift} we plot the scattering phase shift $\delta$ versus $K/K_\text{th}$ for various mass ratios. 
%$\delta$ decreases monotonically from zero as the incident momentum $K$ increases from $0$ to $K_text{th}$.
The minimum of $\delta$ exhibits a strong dependence on the mass ratio $m_1/m_2$.
For larger values of the mass ratio, $\delta$ decreases more rapidly
as $K$ increases initially.
In Fig.~\ref{fig:phaseshift} we also show the approximate results for the phase shifts for a large mass ratio, $m_1/m_2=40$, using the Born-Oppenheimer approximation discussed in Sec.~\ref{sec: BO}.
$\delta$  approaches $-\pi/2$ as $K\to\infty$ for any mass ratio.

We also extracted the dimer-wall scattering length $a_R$ and effective range $r_R$
from the scattering phase shifts at low incident momenta ($K\ll K_\text{th}$)
and plotted the results versus $\ln(m_1/m_2)$ in Fig.~\ref{fig:aRre}. 
Notably, as analytically confirmed by  Eqs. (\ref{eq: delta1}) and (\ref{eq: delta3}),  the dimer-wall scattering length $a_R$ is $a/2$ for the mass ratio $m_1/m_2=1$ and $3a/4$ for $m_1/m_2=3$, and the effective range $r_e$ vanishes for $m_1/m_2=1$ and for $m_1/m_2=3$. 
%For mass ratios 1 and 3, our numerical results for $a_R$ and $r_R$ reduce to the analytical formulas obtained in Sec.~\ref{sec: integrale cases}.
The curves for $a_R$ and $r_R$ approach straight lines at large $\ln(m_1/m_2)$ in Fig.~\ref{fig:aRre}, and
we explain this finding using the Born-Oppenheimer approximation in Sec.~\ref{sec: BO}.

When the incident kinetic energy of the dimer exceeds its binding energy, such that $K>K_\text{th}$, the collision of the dimer with the wall could break up the dimer.  We numerically solved \Eq{integral eq} and determined the reflection coefficient $R=|f|^2$ for various values of the incident momentum and the mass ratio. 
%edit

\begin{figure}
    \centering
    \includegraphics[width=0.85\linewidth]{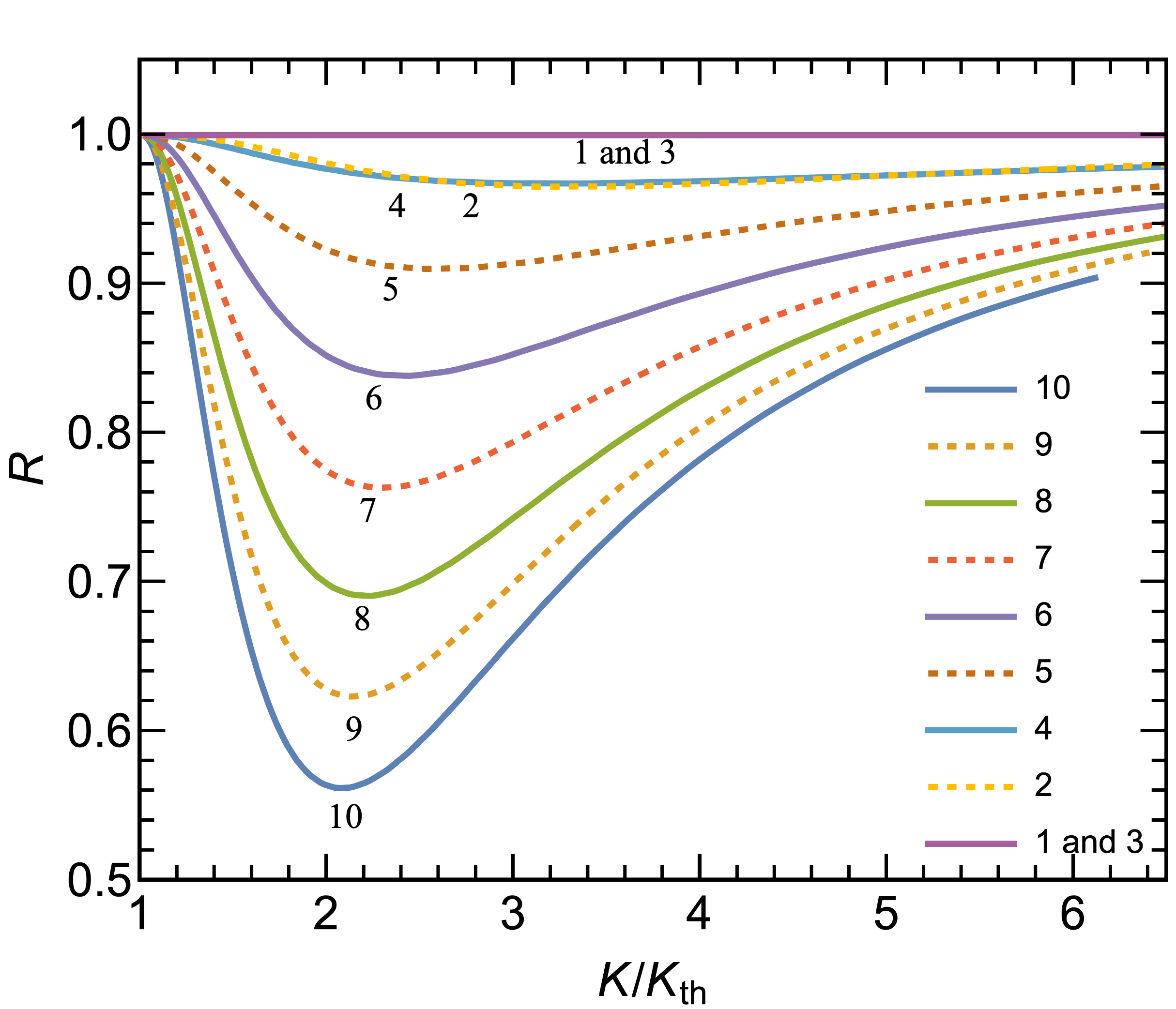}
    \caption{The reflection coefficient $R$ as functions of $K/K_\text{th}$ for mass ratios $m_1/m_2=1$ to $10$. The numbers below the curves are the mass ratios.}
    \label{fig:R1}
\end{figure}

\begin{figure}
    \centering
    \includegraphics[width=0.9\linewidth]{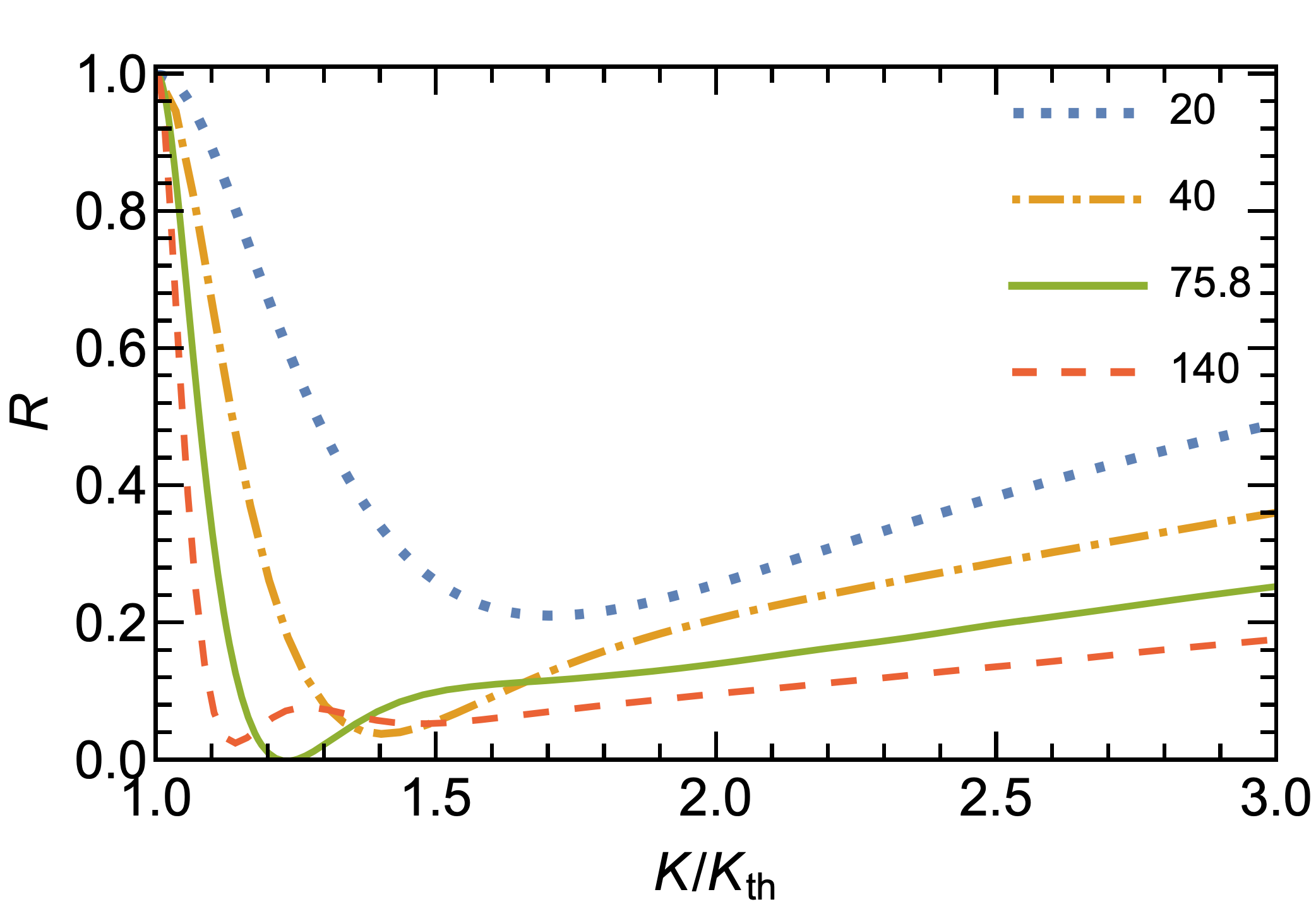}
    \caption{The reflection coefficient $R$ as functions of $K/K_\text{th}$ for mass ratios $m_1/m_2=20$, $40$, $75.8$ and $140$. For $m_1/m_2\approx75.8$, $R$ reaches zero at $K/K_\text{th}\approx 1.24$.}
    \label{fig:R2}
\end{figure}

\begin{figure}
    \centering
    \includegraphics[width=0.9\linewidth]{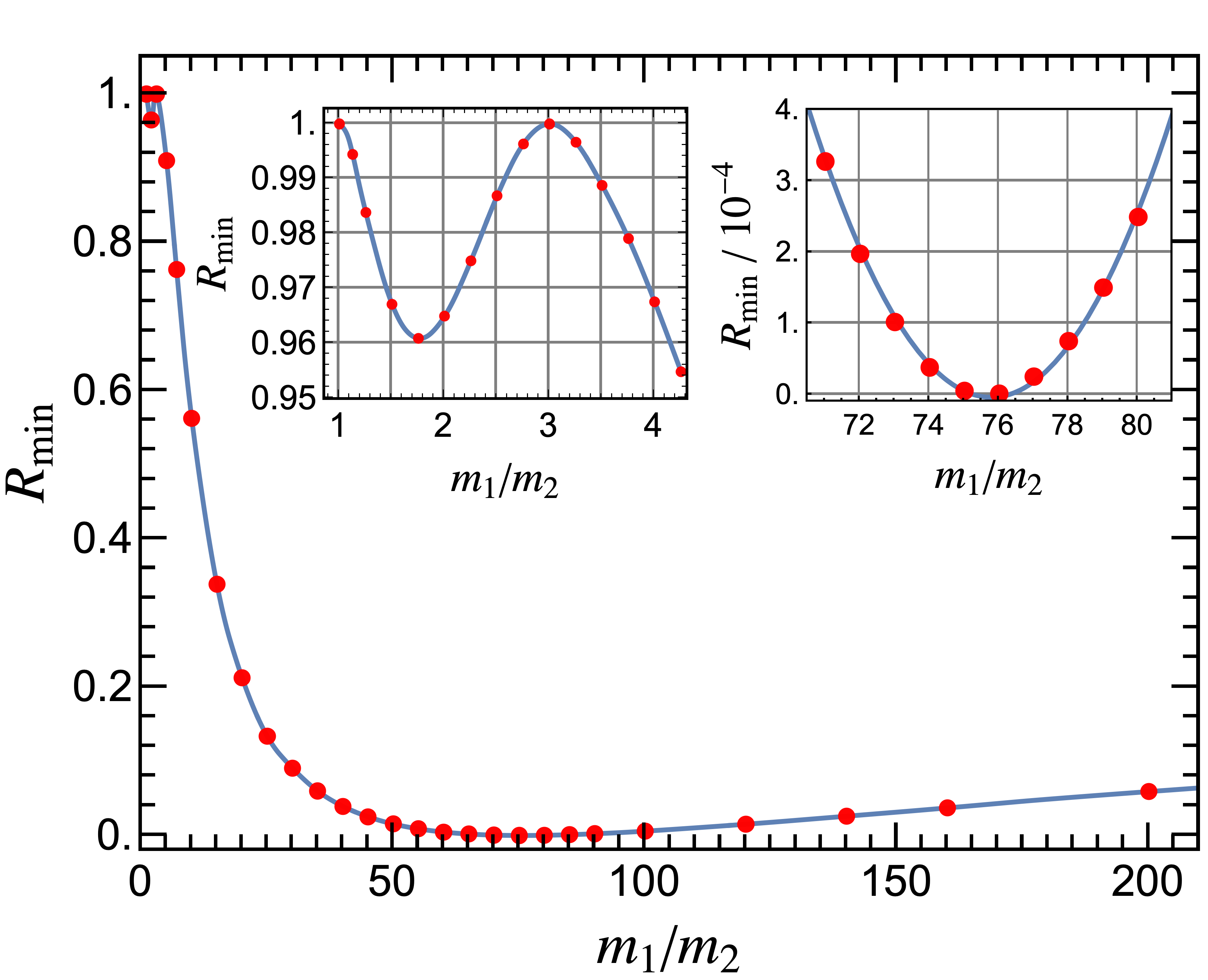}
    \caption{The minimum value of the reflection coefficient $R_{\text{min}}$ for different mass ratios. The left inset plots $R_\text{min}$ for $1\le m_1/m_2\le 4.3$. The right inset 
    plots $R_\text{min}$ for $70.5\le m_1/m_2\le 81$. We numerically found that $R_{\text{min}}=0$ at a critical mass ratio $m_1/m_2\approx75.8$.}
    \label{fig:Rmin}
\end{figure}

We plot $R$ versus  $K/K_{\text{th}}$ in Fig.~\ref{fig:R1} for mass ratios ranging from 1 to 10, and in Fig.~\ref{fig:R2} for large mass ratios 20, 40, 75.8, and 140.
Notably, for mass ratios 1 and 3, the reflection coefficient remains equal to one for all collision energies, indicating full reflection without dissociation, and this is a consequence of integrability  discussed in Sec.~\ref{sec: integrale cases}.
For other mass ratios, $R$ generally first decreases as the collision energy increases, and then increases as the collision energy further increases. Furthermore, systems with larger mass ratios demonstrate a more profound suppression of $R$ for appropriate energies, revealing the effect of mass imbalance on the inelastic scattering process.
In Fig. \ref{fig:Rmin} we plot the minimum value of the reflection coefficient, $R_{\text{min}}$, for different mass ratios.
The two insets in Fig. \ref{fig:Rmin} show the moderate mass ratio regime $1\le m_1/m_2\le4.3$ and the regime where $R_{\text{min}}$ is close to zero, respectively. In the moderate mass ratio regime, $R$ remains close to unity, while a significant mass imbalance substantially enhances the probability of dissociation. And we find that $R_{\text{min}}$ can reach zero at a critical mass ratio of about $75.8$ (at $K/K_\text{th}\approx 1.24$ as shown in \Fig{fig:R2}). As the incident momentum goes to infinity, the reflection coefficient  eventually approaches unity. This  effect can be analytically understood as arising from the asymptotic scaling $R= 1-O(1/K^2)$,  which we will derive in Sec. \ref{sec: high energy}.

\begin{figure}
    \centering
    \subfigure[$m_1/m_2=2$]{
    \includegraphics[width=0.9\linewidth]{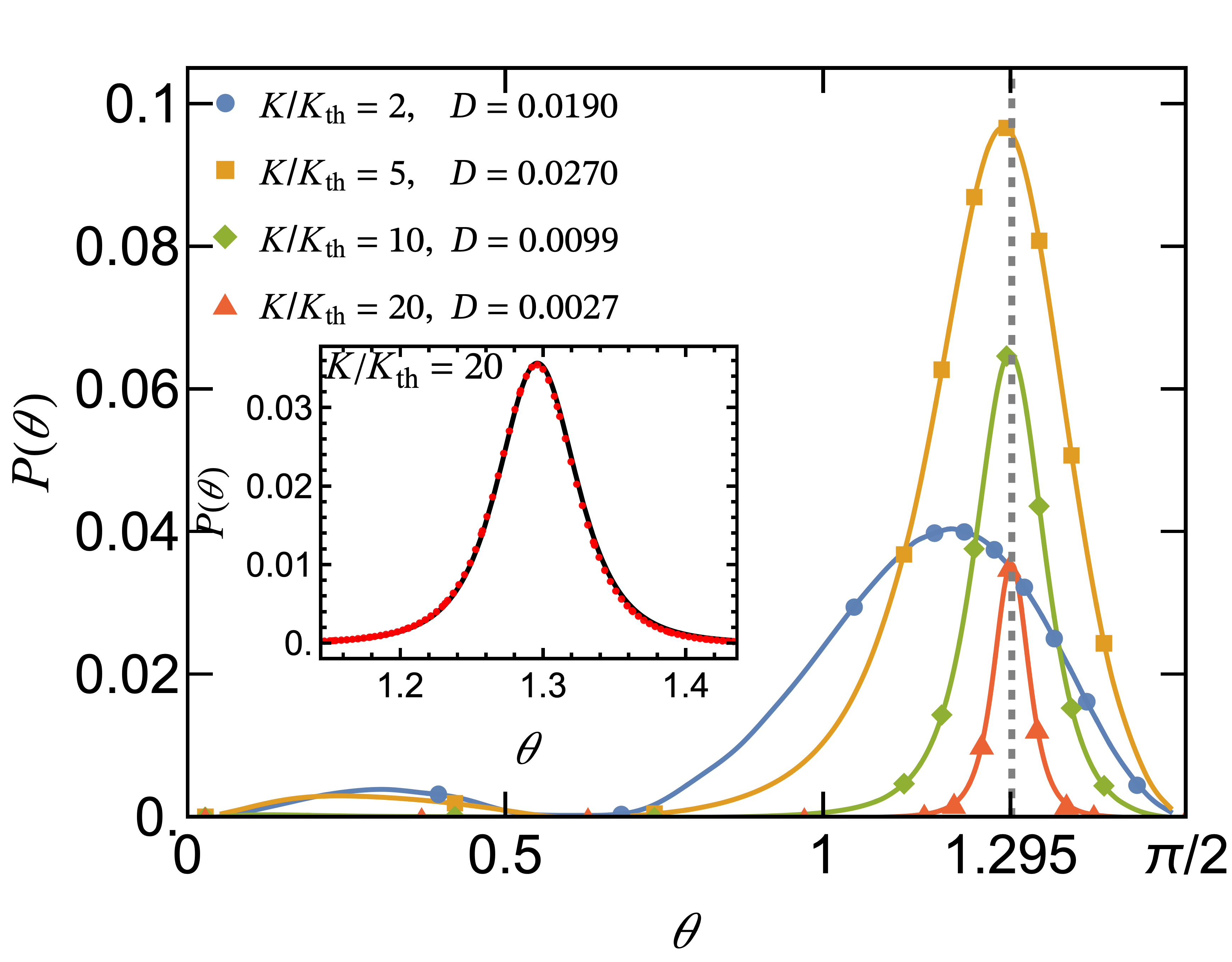}
    }
    \subfigure[$m_1/m_2=4$]{
    \includegraphics[width=0.9\linewidth]{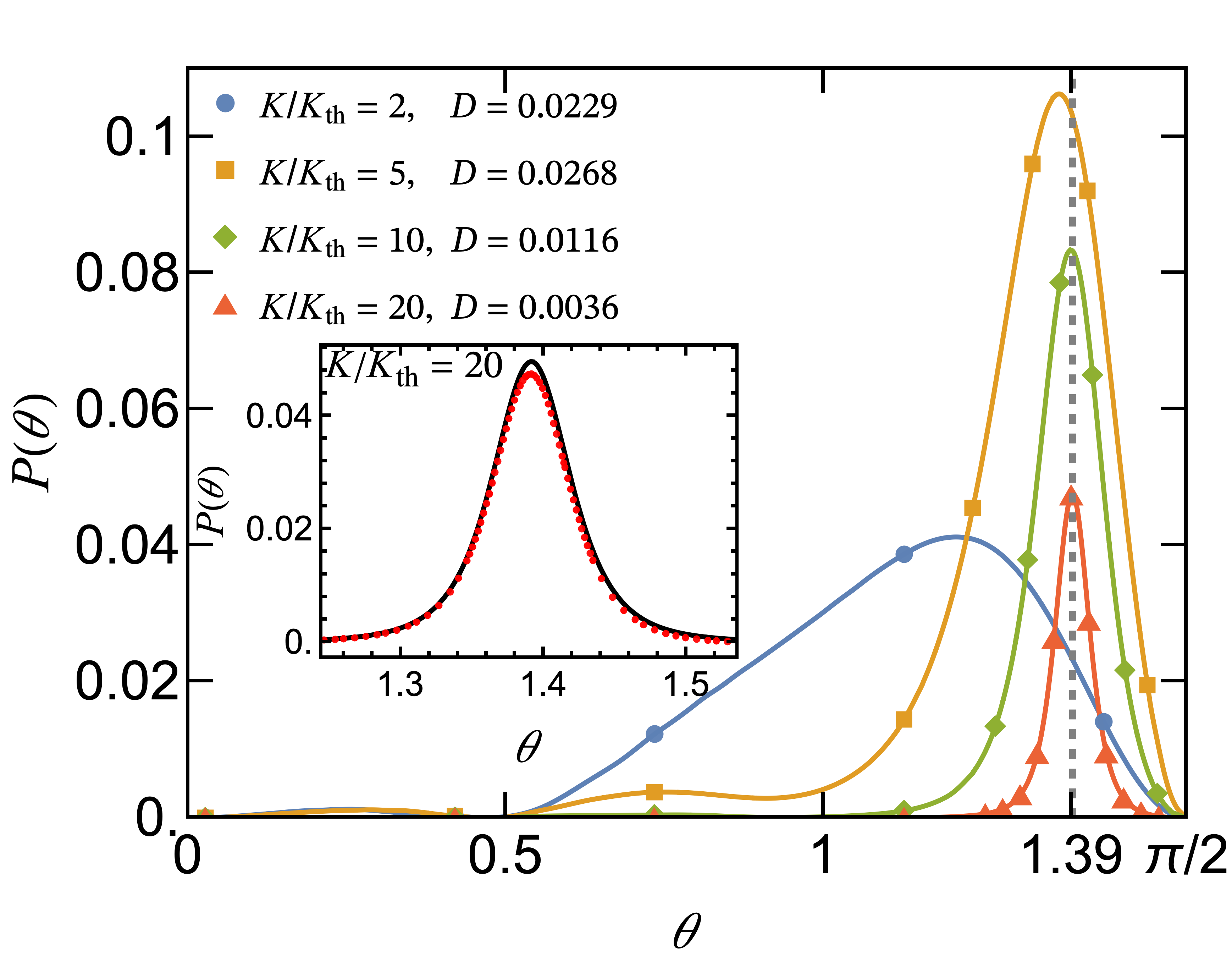}
    }
    \caption{The ``angular distributions" $P(\theta)$ for the dissociated pair for mass ratios $2$ (upper graph) and $4$ (lower graph) at $K/K_\text{th}=$ 2, 5, 10, 20. The corresponding values of the dissociation probability $D$ according to \Eq{D} are shown in the graphs. 
    For larger and larger values of $K/K_\text{th}$,
    the curves approach narrower and narrower peaks centered around $\theta=\theta_c$ where 
$\theta_c$ is defined in \Eq{thetac} and is indicated by a vertical dashed line in each graph.
 The insets compare the semiclassical results according to \Eq{AnalPtheta} (black solid line) 
 with the exact numerical results (red dots) for $K/K_\text{th}=20$.
    }
    \label{fig:Ptheta}
\end{figure}

If the collision breaks up the dimer, the dissociated pair may move along different ``directions" on the $(x_1,x_2)$ plane. The ``direction" is characterized by the ``angle" $\theta$ defined in \Eq{theta}, and in \Fig{fig:Ptheta} we plot the ``angular distribution" $P(\theta)$ for mass ratios $2$ and $4$ at several incident momenta. 
At larger and larger incident momenta, the angular distribution becomes a narrower and narrower peak centered around a particular ``angle" $\theta_c$ whose value depends on the mass ratio. In Sec.~\ref{sec: high energy} we shall derive this feature using a semiclassical analysis.

\section{Born-Oppenheimer Approximation} \label{sec: BO}
For large mass ratios ($m_1/m_2\gg1$) we can use the Born-Oppenheimer (BO) approximation to solve the dimer-wall collision problem. If the heavy particle with mass $m_1$ is fixed at a distance $x>a/2$ from
the wall, the light particle with mass $m_2$ has a bound state with energy $-\kappa^2(x)/2m_2$, where
\begin{equation}\label{BO kappa exact}
\kappa(x)=\frac{1}{a}\left[1+\frac{a}{2x}W\Big(-\frac{2x}{a}e^{-\frac{2x}{a}}\Big)\right],
\end{equation}
and $W$ is the Lambert W function. 
Thus, the heavy particle is governed by an effective single-particle Schrödinger equation: 
\begin{equation}\label{Schrodinger BO}
    -\frac{1}{2 m_1}\frac{d^2\phi(x)}{dx^2}+V_\text{eff}(x)\phi(x)=\frac{ K^2}{2 m_1}\phi(x),
\end{equation}
where $\phi(0)=0$ because of the hard wall, and
\beq\label{V eff}
V_\text{eff}(x)=\frac{1}{2m_2 a^2}-\frac{\kappa^2(x)}{2m_2}
\eeq
is the effective potential experienced by the heavy particle. 
In Fig.~\ref{fig:phaseshift} we show the numerical results for the dimer-wall scattering phase shift $\delta$ based on the BO approximation, Eqs.~\eqref{BO kappa exact}, \eqref{Schrodinger BO} and \eqref{V eff}, for the mass ratio $m_1/m_2=40$ at $K/K_\text{th}\le1$, and find good agreement with the exact numerical results except when $K/K_\text{th}$ is comparable to 1.

At $x\gg a$,
\beq\label{Veff approx}
V_\text{eff}(x)\approx\frac{1}{m_2a^2}e^{-2x/a}.
\eeq
For low-energy collisions ($K\ll K_\text{th}$) the heavy particle is strongly expelled by the barrier potential $V_\text{eff}(x)$, so that $V_\text{eff}(x)$ may be approximated by \Eq{Veff approx}; making this approximation for $V_\text{eff}(x)$, we can solve \Eq{Schrodinger BO} analytically
and find an approximate formula for the dimer-wall scattering phase shift:
\begin{equation}\label{BO delta}
\delta\approx\frac{\ln2}{2}Ka+\arg\Gamma(1+\I K a)-\frac{Ka}{2}\ln\frac{m_1}{m_2},
\end{equation}
where $\Gamma(z)$ is the Gamma function. 
In Fig.~\ref{fig:phaseshift} we show the predictions of \Eq{BO delta} for $m_1/m_2=40$,
and find that they are close to our numerical results for the dimer-wall scattering phase shifts at $K\ll K_\text{th}$. In Fig.~\ref{fig:phaseshift} we see that the predictions based on Eqs.~\eqref{BO kappa exact}, \eqref{Schrodinger BO} and \eqref{V eff} are a better approximation to the exact numerical results for the phase shift than the predictions based on \Eq{BO delta} at $m_1/m_2=40$, but that the two approximations are both very close to the exact numerical results for the phase shift at $K/K_\text{th}\ll1$.

\begin{figure}[htp]
    \centering
    \includegraphics[width=0.85\linewidth]{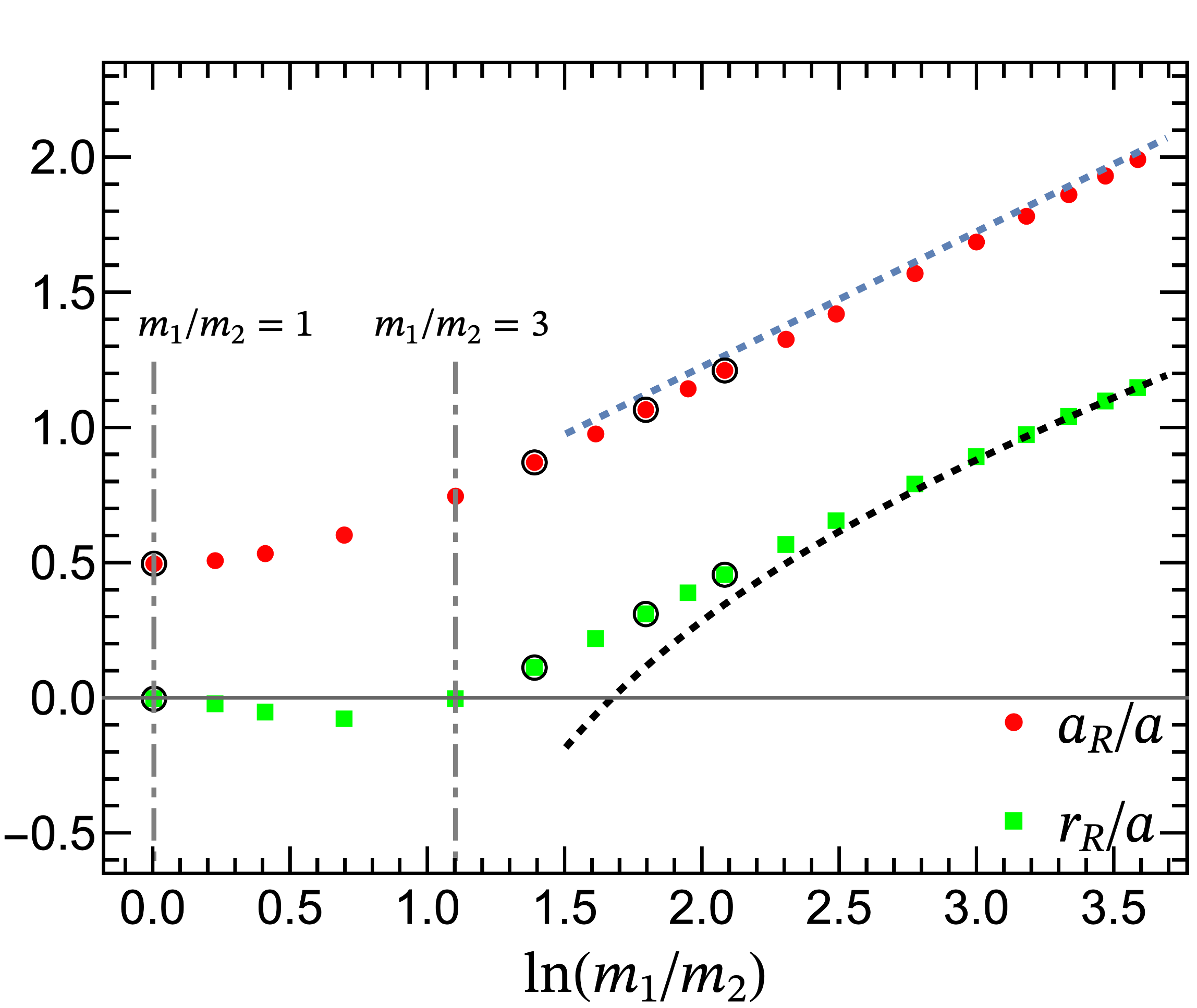}
    \caption{Dimer-wall scattering length $a_R$ and effective range $r_R$ (in units of the two-body scattering length $a$) versus the natural logarithm of the mass ratio. 
    The red dots and the green squares show the values of $a_R/a$ and $r_R/a$, respectively, according to the numerical solution to \Eq{integral eq}. The circles are the predictions of Ref.~\cite{Lee_2011}. The blue dashed line shows the  prediction of \Eq{eq:aR} based on the Born-Oppenheimer (BO) approximation. The black dashed curve shows the prediction of \Eq{eq:re} based on the BO approximation. The vertical dot-dashed lines indicate the integrable cases ($m_1/m_2=1$ or 3), for which there are exact results discussed in Sec.~\ref{sec: integrale cases}.}
    \label{fig:aRre}
\end{figure}

Substituting \Eq{BO delta} into the left hand side of \Eq{kcotdelta} and expanding it in powers
of $K$, we get
\begin{equation}
    a_R\approx\frac{1}{2}\left(\ln\frac{m_1}{m_2}+2\gamma-\ln2\right)a,
    \label{eq:aR}
\end{equation}
\begin{equation}
    r_R\approx\frac{2}{3}a_R\left(1-C\frac{a^3}{a_R^3}\right),  %\qquad C=-\frac{\psi^{(2)}(1)}{2}\approx1.20206,
    \label{eq:re}
\end{equation}
where $\gamma\approx 0.577216$ is Euler's constant, and
$
C=-\frac12\psi^{(2)}(1)\approx1.20206.
$
Here $\psi^{(n)}(\xi)$ is the polygamma function of order $n$.
Our numerical results for $a_R$ and $r_R$ approach the BO formulas in Eqs.~\eqref{eq:aR} and \eqref{eq:re} at large mass ratios, as shown in Fig.~\ref{fig:aRre}.


\section{High energy collisions\label{sec: high energy}}
%To understand the behavior of reflection ($R$) and dissociation($D$) at high energies,  we perform a semi-classical analysis.
For any fixed mass ratio, if $K/K_\text{th}$ is sufficiently large, the de Broglie wave lengths of the incident particles are very short compared to the size of the dimer, and we can understand the problem semiclassically, assuming that particle 1
has incident velocity $v_1$ and particle 2 has incident velocity $v_2$.
Since the magnitude of the center-of-mass velocity of the incident pair is $v_c=K/M$, we have
\begin{equation}
    v_1=-v_c+q/m_1 \qquad v_2=-v_c-q/m_2,
\end{equation}
where $q$ ($-q$) is the momentum of particle 1 (particle 2) in the center-of-mass frame. The normalized wave function of the dimer in the center-of-mass frame is
$\psi_\text{rel}(r)=\frac{1}{\sqrt{a}}\exp(-|r|/a)$. The normalized wave function
in the momentum representation is thus
\begin{equation}\label{psitilderel}
    \widetilde{\psi}_\text{rel}(q) = \frac{1}{\sqrt{2\pi}} \int_{-\infty}^{\infty} \psi_\text{rel}(r) e^{-\I qr} \, dr=\sqrt{\frac{2}{\pi}} \frac{\sqrt{a}}{1+a^2q^2},
\end{equation}
and the probability of finding the relative momentum in the interval $(q,q+dq)$
is $|\widetilde{\psi}_\text{rel}(q)|^2dq$. Since $q\sim 1/a$ according to this probability distribution, we have $v_1\approx v_2\approx -K/M$ at sufficiently large $K/K_\text{th}$.

If particle 1 (the heavier particle) is on the left of particle 2 (namely $x_1<x_2$) before the dimer-wall collision,
particle 1 hits the wall first, and its velocity is changed to $-v_1$ 
after the collision with the wall. Particle 1 then hits particle 2 with relative velocity $v_r\approx 2K/M$, and the probability that they are bounced back from each other is $D_1=1/(1+q'^2a^2)\approx 1/(q'^2a^2)$ according to the $\delta$ function interaction potential, where $q'=\mu v_r\approx 2m_1m_2K/M^2$ is the momentum of particle 1 in the new center-of-mass frame. So $D_1\approx M^2K_\text{th}^2/(4m_1m_2K^2)$. After the two particles bounce back from each other, particle 1 has a new velocity $v_1'$
and particle 2 has a new velocity $v_2'$ satisfying momentum conservation
\beq
m_1v_1'+m_2v_2'=-m_1v_1+m_2v_2
\eeq
and energy conservation
\beq
\frac{1}{2}m_1v_1'^2+\frac{1}{2}m_2v_2'^2=\frac{1}{2}m_1v_1^2+\frac{1}{2}m_2v_2^2.
\eeq
Solving the above equations, we get
\begin{equation}
    v_1'=\frac{m_1-3m_2}{m_1+m_2}v_c+\frac{(-3m_1+m_2)q}{m_1(m_1+m_2)},
\end{equation}
\begin{equation}
    v_2'=\frac{3m_1-m_2}{m_1+m_2}v_c+\frac{(m_1-3m_2)q}{m_2(m_1+m_2)}.
\end{equation}
If $m_1/m_2<3$, then for sufficiently large $v_c$, $v_1'<0$, and particle 1 will hit the wall again and its velocity changes to $-v_1'$. 
In the distant future after the collisions, the coordinates of the particles have a ratio $x_2/x_1\approx |v_2'|/|v_1'|$ regardless of the mass ratio. Therefore, according to \Eq{theta} we have
\beq
\theta=\arctan\Big(\sqrt{\frac{m_2}{m_1}}\frac{|v_2'|}{|v_1'|}\Big).
\eeq
At large $v_c$ we may Taylor expand $\theta$ to first order in $q$ to find
\beq \label{theta q relation}
\theta\approx\theta_c+\frac{\sigma}{v_c\sqrt{m_1m_2}}q,
\eeq
where $\sigma=+1$ if $m_1>3m_2$ and $\sigma=-1$ if $m_1<3m_2$, and
\begin{equation}
    \theta_c=\arctan\Big(\Big|\frac{3m_1-m_2}{3m_2-m_1}\Big|\sqrt{\frac{m_2}{m_1}}\Big).
    \label{thetac}
\end{equation}
Therefore, if the dimer breaks up after collision with the wall,
the ``angle" $\theta$ in the distant future has a probability density of approximately
$v_c\sqrt{m_1m_2}|\widetilde{\psi}_\text{rel}(q)|^2$, and we find
\beq \label{AnalD}
D\approx D_1\approx\frac{(m_1+m_2)^2}{4m_1m_2}\frac{K_\text{th}^2}{K^2},
\eeq
and
\beq
P(\theta)\approx D_1v_c\sqrt{m_1m_2}|\widetilde{\psi}_\text{rel}(q)|^2.
\eeq
Combining Eqs.~\eqref{psitilderel}, \eqref{theta q relation} and \eqref{AnalD} with the above equation, we find
\beq\label{AnalPtheta}
P(\theta)\approx \frac{(m_1+m_2)^2 K_\text{th}}{2\pi m_1 m_2 K\big[1+(K/K_\text{th})^2(\theta-\theta_c)^2\big]^2}.
\eeq

If particle 1 (the heavier particle) is on the right of
particle 2 before the dimer-wall collision, particle 2 hits the wall first, and its velocity is changed to $-v_2$. Then there is a small probability $D_1$ that particle 2 is bounced off by particle 1. Then particle 2 hits the wall again, and its velocity gets reversed. Then
\begin{itemize}
\item if $m_1>3m_2$ particle 2 has nearly 100\% probability of passing through particle 1, and then particle 1 hits the wall, and its velocity is reversed;
\item if $m_1<3m_2$ particle 2 catches up with particle 1 and has nearly 100\% probability of passing through particle 1.
\end{itemize}
One can similarly show that in this case the approximate formulas for $D$ and $P(\theta)$ in Eqs.~\eqref{AnalD} and \eqref{AnalPtheta} remain valid.

Equation~\eqref{AnalPtheta} shows that $P(\theta)$ peaks at $\theta\approx\theta_c$ and the peak has a narrow width at $K/K_\text{th}\gg1$.
In the insets of \Fig{fig:Ptheta}, we compare the predictions of \Eq{AnalPtheta}
with our numerical results for $P(\theta)$ based on Eqs.~\eqref{integral eq}, \eqref{c}, and \eq{Ptheta}, and find excellent agreement when $K/K_\text{th}$ is large.

\begin{figure}[htp]
    % \vspace{0.5cm}
    \centering
    \includegraphics[width=0.85\linewidth]{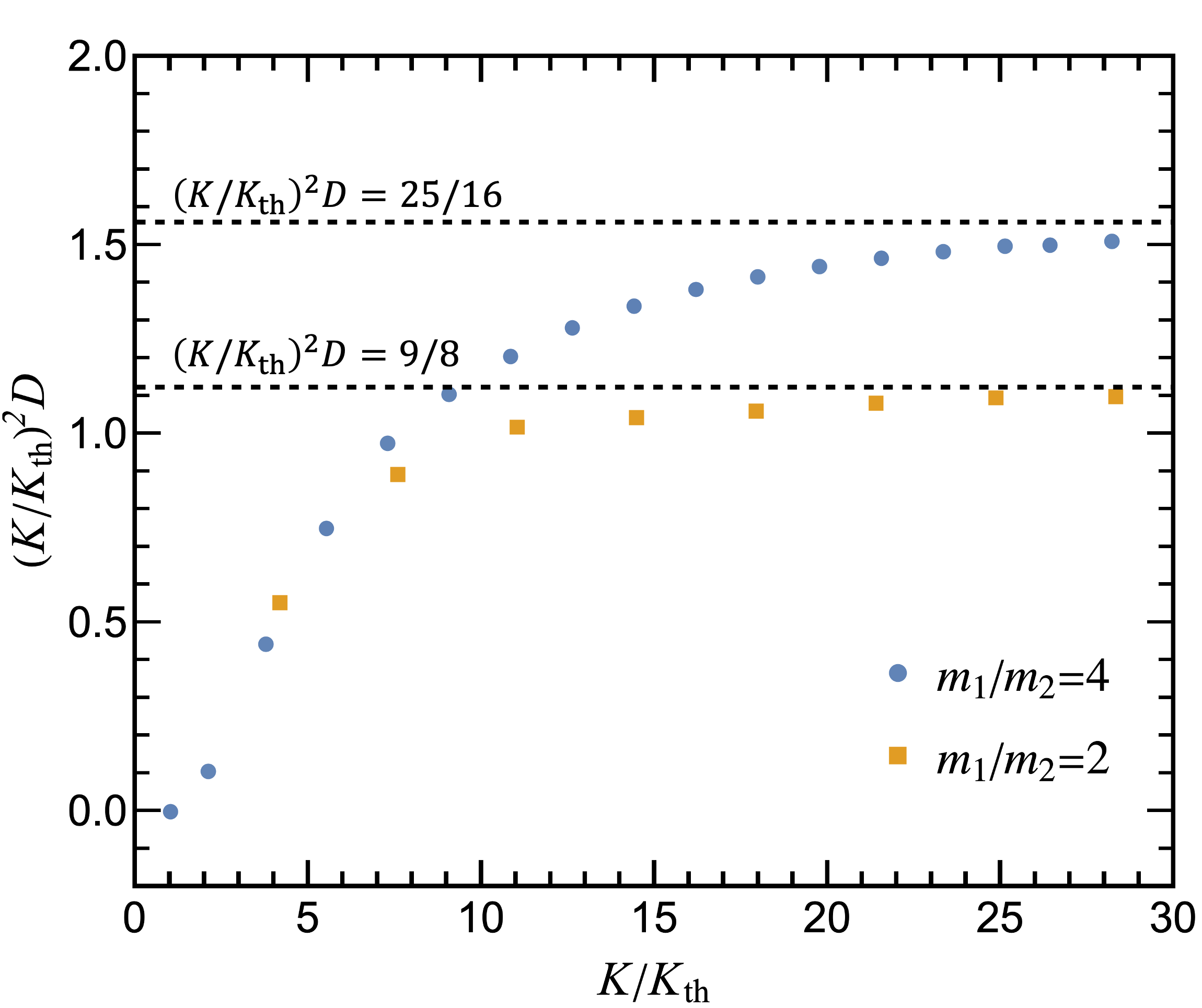}
    \caption{The scaled dissociation probability $(K/K_\text{th})^2D$ as functions of $K/K_\text{th}$ for the mass ratios $2$ and $4$. In the high energy limit, $(K/K_\text{th})^2D$ approaches a constant equal to $\frac{(m_1+m_2)^2}{4m_1m_2}$ according to Eq. (\ref{AnalD}).}
    \label{fig:D}
\end{figure}

In \Fig{fig:D} we plot $(K/K_\text{th})^2D$ [where $D=1-R=1-|f|^2$ and $f$ is obtained by numerically solving Eqs.~\eqref{integral eq} and \eqref{psi large x}] as functions of $K/K_\text{th}$ for the mass ratios $2$ and $4$, and numerically verify the validity of \Eq{AnalD} at large $K/K_\text{th}$.

\section{Summary}
In this work, we have studied the scattering of a dimer from a hard wall in one dimension, assuming a delta function attractive interaction potential between the two constituent particles. 
We have numerically computed the scattering phase shift and the reflection coefficient as functions of the mass ratio and the collision energy by solving a one-dimensional integral equation.
For large mass ratios we used the Born-Oppenheimer approximation to find formulas for the elastic scattering phase shift, the dimer-wall scattering length, and the dimer-wall effective range.
For high energy collisions, we used a semiclassical analysis to derive approximate formulas for the dissociation probability and the ``angular distribution" of the dissociated pair.
For the mass ratio of about $75.8$, we find that the reflection coefficient vanishes at a particular collision energy.
For the mass ratios of 1 and 3, the problem is integrable and the reflection coefficient is equal to unity for all collision energies.

\begin{acknowledgments}
This work was supported by the National Key R$\&$D Program of China Grant No. 2021YFA1400902,
the National Natural Science Foundation of China Grant No. 92365202,
and the National Key R$\&$D Program of China Grant No. 2019YFA0308403.
\end{acknowledgments}

\bibliography{ref}

\end{document}